\documentclass[epj-spec]{svjour}

\usepackage{graphics}
\usepackage{graphicx}
\usepackage{epsfig}

\begin{document}

\title{Detecting chaos, determining the dimensions of tori and predicting slow
diffusion in Fermi--Pasta--Ulam lattices by the Generalized Alignment Index
method}

\author{Charalampos Skokos\inst{1}\fnmsep\thanks{\email{hskokos@imcce.fr}}
\and Tassos Bountis\inst{2}\fnmsep\thanks{\email{bountis@math.upatras.gr}}
\and Chris
Antonopoulos\inst{2}\fnmsep\thanks{\email{antonop@math.upatras.gr}}}

\institute{Astronomie et Syst\`{e}mes Dynamiques, IMCCE, Observatoire de
Paris, 77 avenue Denfert--Rochereau, F-75014, Paris, France \and Department of
Mathematics, Division of Applied Analysis and Center for Research and
Applications of Nonlinear Systems (CRANS), University of Patras, GR-26500,
Patras, Greece}

\abstract{The recently introduced GALI method is used for rapidly detecting
chaos, determining the dimensionality of regular motion and predicting slow
diffusion in multi--dimensional Hamiltonian systems. We propose an efficient
computation of the GALI$_k$ indices, which represent volume elements of $k$
randomly chosen deviation vectors from a given orbit, based on the Singular
Value Decomposition (SVD) algorithm. We obtain theoretically and verify
numerically asymptotic estimates of GALIs long--time behavior in the case of
regular orbits lying on low--dimensional tori. The GALI$_k$ indices are
applied to rapidly detect chaotic oscillations, identify low--dimensional tori
of Fermi--Pasta--Ulam (FPU) lattices at low energies and predict weak
diffusion away from quasiperiodic motion, long before it is actually observed
in the oscillations.}

\maketitle

%-------------------------------------------------
\section{Introduction}
\label{intro}

A great variety of physical systems can be described by Hamiltonian systems or
symplectic maps \cite{Chirikov,MacKay_1987,LL92}. Their applications range
from the stability of the solar system \cite{Contopoulos} and the containment
of charged particles in high intensity magnetic fields \cite{LL92} to the
blow--up of hadron beams in high energy accelerators \cite{Scandale} and the
study of simple molecules and hydrogen--bonded systems
\cite{Farantos,Jung_06}. More recently, much attention has been focused on the
dynamics of nonlinear lattices and in particular on localization in the form
of q--breathers \cite{Flach_05} energy transport and equipartition properties
of Fermi--Pasta--Ulam (FPU) particle chains \cite{Ford}. The main difficulty
with determining the nature of the motion in Hamiltonian dynamics is that
regular and chaotic orbits are distributed in phase space in very intricate
ways, in contrast with dissipative systems, where all orbits eventually fall
on regular or chaotic attractors.

The most widely used method for distinguishing order from chaos in dynamical
systems is the evaluation of the Lyapunov Exponents (LEs) $\sigma_i$ for each
given orbit. Benettin {\it et al.}  \cite{BGGS80a} studied the problem of
determining all LEs theoretically and proposed in \cite{BGGS80b} an algorithm
for their numerical computation. In practice, one often calculates only the
largest LE, $\sigma_1$, following one deviation vector from the given orbit
and if $\sigma_1>0$ the orbit is characterized as chaotic.

In the present paper, we apply the Generalized ALignment Index of order $k$,
GALI$_k$, $k=2,\ldots,2N$ \cite{SBA_07}, to distinguish efficiently between
regular and chaotic orbits of multi--dimensional Hamiltonian systems of $N$
degrees of freedom. These indices generalize a similar indicator called
Smaller ALignment Index (SALI) \cite{S01}, in that they use information of
{\it more than two} deviation vectors from the reference orbit (see
\cite{SBA_07} for more details). In particular, the GALI$_k$ is proportional
to volume elements formed by $k$ initially linearly independent unit deviation
vectors whose magnitude is normalized to unity at every time step.

After recalling briefly the definition of GALI$_k$ and its general behavior
for regular and chaotic motion in Sect.~\ref{GALI-def}, we propose in
Sect.~\ref{GALI-num} a technique for the efficient computation of GALI$_k$,
based on the Singular Value Decomposition (SVD) of the matrix having as rows
the $k$ normalized deviation vectors. In Sect.~\ref{GALI-low} we study
theoretically the behavior of GALIs for regular orbits of $N$ degrees of
freedom Hamiltonian systems that lie on an $s$--dimensional torus, with $1\leq
s\leq N$.  In Sect.~\ref{Apl} we present applications of the GALI$_k$ approach
to an $N=8$ particles FPU lattice and show that the GALI$_k$ indices: (i) can
be used for the rapid discrimination between regular and chaotic motion, (ii)
determine the correct dimensionality of tori and (iii) predict that the motion
is weakly chaotic, long before this can be seen in the actual oscillations.

By `weak chaos' we mean weakly diffusive motion through a network of
resonances. On the other hand, when one speaks of `sticky' orbits
\cite{Sticky}, one generally refers to motion occurring just outside the
boundary of a large regular region, where orbits remain for a long times
before rapidly escaping to distant parts of phase space through a large
chaotic sea. In our case, `weakly chaotic' motion occurs within regimes
`surrounded' by (often high--dimensional) tori, is characterized by very small
LEs and is reminiscent of what is called in the literature Arnol'd diffusion
\cite{LL92}. Finally, in Sect.~\ref{sum}, we summarize the results and present
our conclusions.

%------------------------------------------------------
\section{The GALI method}
\label{GALI}

%------------------------------------------------------
\subsection{Definition and behavior of GALI}
\label{GALI-def}

Following \cite{SBA_07} let us first briefly recall the definition of GALI and
its behavior for regular and chaotic motion.  We consider a Hamiltonian system
of $N$ degrees of freedom having a Hamiltonian $H(q_1, \ldots, q_N,p_1,
\ldots, p_N)$ where $q_i$ and $p_i$, $i=1,\ldots,N$ are the generalized
coordinates and momenta respectively. An orbit of this system is defined by a
vector $\vec{x}(t)=(x_1(t), \ldots, x_{2N}(t))$, with $x_i=q_i$,
$x_{i+N}=p_i$, $i=1,\ldots,N$. This orbit is a solution of Hamilton's
equations $d \vec{x}/dt= \vec{\mathcal{V}}(\vec{x})= \left( \partial
H/\partial \vec{p}\, , - \partial H/\partial \vec{q} \right)$, while the
evolution of a deviation vector $\vec{w}(t)$ from $\vec{x}(t)$ obeys the
variational equations $d \vec{w}/dt = \textbf{M}(\vec{x}(t)) \,\vec{w}$, where
$\textbf{M}= \partial \vec{\mathcal{V}} / \partial \vec{x}$ is the Jacobian
matrix of $\vec{\mathcal{V}}$.

Let us follow $k$ normalized deviation vectors $\hat{w}_1$, $\ldots$,
$\hat{w}_k$ (with $2\leq k \leq 2N$) in time, and determine whether they
become linearly dependent, by checking if the volume of the corresponding
$k$--parallelogram goes to zero. This volume is equal to the norm of the wedge
or exterior product \cite{Spivak_1999} of these vectors. Hence we are led to
define the following `volume' element:
\begin{equation}
\mbox{GALI}_k(t)=\| \hat{w}_1(t)\wedge \hat{w}_2(t)\wedge \cdots
\wedge\hat{w}_k(t) \|\,\, , \label{eq:GALI}
\end{equation}
which we call the Generalized Alignment Index (GALI) of order $k$.  We note
that the hat ($^\wedge $) over a vector denotes that it is of unit magnitude.
Thus, for each initial condition $\vec{x}(0)$, we solve Hamilton's equations
for $\vec{x}(t)$, together with their variational equations for $k$ initially
linearly independent deviation vectors $\hat{w}_i$, $i=1,\ldots,k$. Clearly,
if at least two of these vectors become linearly dependent, the wedge product
in (\ref{eq:GALI}) becomes zero and the volume element vanishes.

In the case of a chaotic orbit all deviation vectors tend to become linearly
dependent, aligning in the direction of the eigenvector which corresponds to
the maximal Lyapunov exponent and GALI$_{k}$ tends to zero exponentially
following the law \cite{SBA_07}:
\begin{equation}
\mbox{GALI}_k(t) \sim e^{-\left[ (\sigma_1-\sigma_2) + (\sigma_1-\sigma_3)+
\cdots+ (\sigma_1-\sigma_k)\right]t},
\label{eq:GALI_chaos}
\end{equation}
where $\sigma_1, \ldots, \sigma_k$ are approximations of the first $k$ largest
Lyapunov exponents. In the case of regular motion on the other hand, all
deviation vectors tend to fall on the $N$--dimensional tangent space of the
torus on which the motion lies. Thus, if we start with $k\leq N$ general
deviation vectors they will remain linearly independent on the
$N$--dimensional tangent space of the torus, since there is no particular
reason for them to become aligned. As a consequence GALI$_{k}$ remains
practically constant for $k\leq N$. On the other hand, GALI$_{k}$ tends to
zero for $k>N$, since some deviation vectors will eventually become linearly
dependent, following a particular power law which depends on the
dimensionality $N$ of the torus and the number $k$ of deviation vectors. So,
the generic behavior of GALI$_k$ for regular orbits lying on $N$--dimensional
tori is given by \cite{SBA_07}:
\begin{equation}
\mbox{GALI}_k (t) \sim \left\{ \begin{array}{ll} \mbox{constant} & \mbox{if
$2\leq k \leq N$} \\ \frac{1}{t^{2(k-N)}} & \mbox{if $N< k \leq 2N$} \\
\end{array}\right. .
\label{eq:GALI_order_all_N}
\end{equation}

%------------------------------------------------------
\subsection{Numerical computation of GALI}
\label{GALI-num}

In order to numerically compute GALI$_k$ in \cite{SBA_07} using
(\ref{eq:GALI}) we considered as a basis of the $2N$--dimensional tangent
space of the Hamiltonian flow the usual set of orthonormal vectors $\hat{e}_1=
(1,0,0,\ldots,0)$, $\hat{e}_2= (0,1,0,\ldots,0)$, ..., $\hat{e}_{2N}=
(0,0,0,\ldots,1)$. So, any unitary deviation vector $\hat{w}_i$ can be written
as:
\begin{equation}
\hat{w}_i=\sum_{j=1}^{2N} w_{ij} \hat{e}_j \,\, ,\,\, i=1,2,\ldots,k
\label{eq:w}
\end{equation}
where $w_{ij}$ are real numbers satisfying $\sum_{j=1}^{2N} w_{ij}^2
=1$. Considering the $k \times 2N$ matrix $\textbf{A}$ having as rows the
coordinates of $k$ such vectors, we can write equations (\ref{eq:w}) in matrix
form as:
\begin{equation}
\left[\begin{array}{c} \hat{w}_1 \\ \hat{w}_2 \\ \vdots \\ \hat{w}_k
 \end{array} \right] = \left[
\begin{array}{cccc}
w_{11} & w_{12} & \cdots & w_{1\, 2N} \\ w_{21} & w_{22} & \cdots & w_{2\, 2N}
\\ \vdots & \vdots & & \vdots \\ w_{k1} & w_{k2} & \cdots & w_{k\, 2N}
\end{array} \right] \cdot \left[\begin{array}{c} \hat{e}_1 \\ \hat{e}_2 \\
\vdots \\ \hat{e}_{2N} \end{array} \right] = \textbf{A} \cdot
\left[\begin{array}{c} \hat{e}_1 \\ \hat{e}_2 \\ \vdots \\ \hat{e}_{2N}
\end{array} \right].\,\,\, \label{eq:matrix_a}
\end{equation}
The norm of the wedge product of the $k$ deviation vectors appearing in
(\ref{eq:GALI}) was defined in \cite{SBA_07} as the square root of the sum of
the squares of the determinants of all possible $k \times k$ submatrices of
matrix $\textbf{A}$. So, for the computation of GALI$_k$ we have:
\begin{equation}
\mbox{GALI}_k =\|\hat{w}_1\wedge \hat{w}_2\wedge \cdots \wedge\hat{w}_k \|=
\left\{\sum_{1 \leq i_1 < i_2 < \cdots < i_k \leq 2N} \left( \det \left[
\begin{array}{cccc}
w_{1 i_1} & w_{1 i_2} & \cdots & w_{1 i_k} \\ w_{2 i_1} & w_{2 i_2} & \cdots &
w_{2 i_k} \\ \vdots & \vdots & & \vdots \\ w_{k i_1} & w_{k i_2} & \cdots &
w_{k i_k} \end{array} \right] \right)^2 \right\}^{1/2} ,\label{eq:norm}
\end{equation}
where the sum is performed over all possible combinations of $k$ indices out
of $2N$.

Eq.~(\ref{eq:norm}) is ideal for the theoretical determination of the
asymptotic (long time) behavior of GALIs for chaotic and regular orbits (see
\cite{SBA_07} for more details), as well as, for regular orbits that lie on
low dimensional tori (see Sect.~\ref{GALI-low} below).  However, from a
practical point of view the computation of determinants is not the most
efficient way of computing GALI$_k$, as has already pointed out in
\cite{AThesis}. For low dimensional systems, the number of determinants
appearing in (\ref{eq:norm}) for the computation of GALI$_k$ is not
prohibitive, but as the number of degrees of freedom increases the computation
can become extremely time consuming and in some cases impractical. For
example, in the case of an $N=8$ degree of freedom system, like the one
studied in Sect.~\ref{Apl}, the computation of GALI$_8$ requires the
evaluation of $12870$ $8\times 8$ determinants, while, GALI$_{15}$ in an
$N=15$ degree of freedom system requires the computation of $155117520$
$15\times 15$ determinants!

We have already mentioned that GALI$_k$ measures the volume of the
$k$--parallelogram $P_k$ having as edges the $k$ unitary deviation vectors
$\hat{w}_i$, $i=1,\ldots, k $ of (\ref{eq:w}) and (\ref{eq:matrix_a})
above. The volume of $P_k$ is then given by (see for instance \cite{HH}):
\begin{equation}
\mbox{vol}(P_k)=\sqrt{\det(\textbf{A} \cdot \textbf{A}^{\mathrm{T}})}\,\,,
\label{eq:vol_k}
\end{equation}
where $(^{\mathrm{T}})$ denotes transpose. Since $\det(\textbf{A} \cdot
\textbf{A}^{\mathrm{T}})$ is equal to the sum appearing in (\ref{eq:norm})
(this equality is called Lagrange's identity, see for instance \cite{B58}), we
have:
\begin{equation}
\mbox{GALI}_k = \sqrt{\det(\textbf{A} \cdot
\textbf{A}^{\mathrm{T}})},\label{eq:gali_det}
\end{equation}
as an alternative way of computing GALI$_k$, where only the multiplication of
two matrices and the square root of one determinant appears.

A different way of evaluating GALI$_k$, which actually proved to be more
accurate, is obtained by performing the Singular Value Decomposition (SVD) of
$\textbf{A}^{\mathrm{T}}$. So, the $2N \times k$ matrix
$\textbf{A}^{\mathrm{T}}$ can be written as the product of a $2N \times k$
column--orthogonal matrix $\textbf{U}$, a $k \times k$ diagonal matrix
$\textbf{Z}$ with positive or zero elements $z_i$, $i=1,\ldots, k$ (the
so--called {\it singular values}), and the transpose of a $k \times k$
orthogonal matrix $\textbf{V}$:
\begin{equation}
\textbf{A}^{\mathrm{T}}=\textbf{U}\cdot \textbf{Z} \cdot
\textbf{V}^{\mathrm{T}}. \label{eq:SVD}
\end{equation}
We note that matrices $\textbf{U}$ and $\textbf{V}$ are orthogonal so
that:
\begin{equation}
\textbf{U}^{\mathrm{T}}\cdot \textbf{U}=\textbf{V}^{\mathrm{T}}\cdot
\textbf{V} = \mathrm{I}_k , \label{eq:UV}
\end{equation}
with $\mathrm{I}_k$ being the $k \times k$ unit matrix. For a more detailed
description of the SVD method, as well as an algorithm for its implementation
the reader is referred to \cite{NumRec} and references therein.  Using
Eq.~(\ref{eq:gali_det}) for the computation of GALI$_k$, as well as
Eqs.~(\ref{eq:SVD}) and (\ref{eq:UV}), we get:
\begin{eqnarray}
\mbox{GALI}_k = \sqrt{\det\left(\textbf{A} \cdot
\textbf{A}^{\mathrm{T}}\right)} = \sqrt{\det \left( \textbf{V}
\cdot\textbf{Z}^{\mathrm{T}} \cdot \textbf{U}^{\mathrm{T}} \cdot \textbf{U}
\cdot \textbf{Z} \cdot \textbf{V}^{\mathrm{T}}\right)} \nonumber\\ = \sqrt{
\det\left( \textbf{V} \cdot \mbox{diag} (z_i^2) \cdot
\textbf{V}^{\mathrm{T}}\right)}= \sqrt{ \det\left( \mbox{diag}
(z_i^2)\right)}=\prod_{i=1}^k z_i \, .\nonumber\
\end{eqnarray}
Thus, we conclude that GALI$_k$ is equal to the product of the singular values
of matrix $\textbf{A}$ (\ref{eq:matrix_a}) defined by the $k$ normalized
deviation vectors and at the same time, we theoretically explain the
computationally verified equality of this product to values of GALI$_k$
computed by (\ref{eq:norm}) which was reported in \cite{AThesis}. The SVD
approach, therefore, provides a very accurate determination of the logarithm
of GALI$_k$, which can now be used for the discrimination between regular and
chaotic motion, as it leads to:
\begin{equation}
\log (\mbox{GALI}_k) =\sum_{i=1}^k \log(z_i) \, ,
\label{eq:gali_svd}
\end{equation}
and for this reason we implement this approach for the computation of GALI$_k$
in numerical applications. From Eq.~(\ref{eq:gali_svd}) we see that the
computation of the singular values $z_i$ with the usual double precision
accuracy, permits the accurate determination of large negative values of $\log
(\mbox{GALI}_k)$, which correspond to very small values of GALI$_k$.

The problem of the numerical computation of orbits, keeping constant
the numerical value of the Hamiltonian $H$, is of great importance
in numerical studies of dynamical systems. Several integration
schemes have been developed and applied over the years with varying
degrees of success (see \cite{MQ06} for a survey of such methods).
In our study we use an 8th order Runge--Kutta method proposed in
\cite{PD81}, both for the numerical integration of Hamilton's
equations (evolution of an orbit), as well as for the variational
equations (evolution of deviation vectors). The scheme proved to be
very efficient and accurate since, in all our computations, we
always kept the relative error of the values of the Hamiltonian
function ($|H(t)-H(0)|/|H(0)|$) below $10^{-12}$.

%------------------------------------------------------
\subsection{Behavior of GALI for regular orbits of low dimensional tori}
\label{GALI-low}

Now we turn to the GALI$_k$ method for regular orbits of an $N$ degree of
freedom Hamiltonian, that lie on $s$--dimensional tori, with $1\leq s \leq
N$. In this case, one could perform a local transformation to action--angle
variables, ${J_i,\theta_i}$, whence Hamilton's equations of motion can be
easily integrated to give $J_i(t) = J_{i0}$, $\theta_i(t) = \theta_{i0}
+\omega_i(J_{10}, \ldots, J_{s0})\, t$, where $J_{i0}$, $\theta_{i0}$ are the
initial conditions ($i=1,\ldots,N$) and $\omega_i \equiv 0$ for $s < i \leq
N$. Denoting by $\xi_i$, $\eta_i$ small deviations from the $J_i$ and
$\theta_i$ respectively and using as basis of the $2N$--dimensional tangent
space of the Hamiltonian flow the $2N$ unit vectors
$\{\hat{v}_1,\ldots,\hat{v}_{2N}\}$, such that the first $N$ of them
correspond to the $N$ action variables and the remaining ones to the $N$
conjugate angle variables, any deviation vector $\vec{w}_i$, $i=1,2,\ldots$
can be written as
\begin{equation}
\vec{w}_i(t) = \sum_{j=1}^{N} \xi_j^i(0) \, \hat{v}_j + \sum_{j=1}^{N}
\left(\eta_j^i(0)+ \sum_{k=1}^N \omega_{jk} \xi_k^i(0) t\right) \hat{v}_{N+j},
\label{eq:order_dev_vec}
\end{equation}
with $\omega_{kj}=\partial \omega_k/ \partial J_j$ computed from the initial
values $J_{j0}$ for $k,j=1,2,\ldots, s$ and $\omega_{kj}\equiv 0$ for
$k,j=s+1,s+2,\ldots, N$.  It follows easily from the above that for
sufficiently long times, $\|\vec{w}(t)\|\sim t$.

Let us now study the case of $k$, initially linearly independent, randomly
chosen, unit deviation vectors $\{\hat{w}_1,\ldots,\hat{w}_{k}\}$ expressed in
terms of the new basis by the transformation $[\begin{array}{ccc} \hat{w}_1
&\ldots &\hat{w}_{k}
\end{array}]^{\mathrm{T}} = \textbf{D}\cdot[
\begin{array}{ccc} \hat{v}_1 &\ldots &\hat{v}_{2N}
\end{array}]^{\mathrm{T}}$. The random choice of the initial deviation vectors
corresponds to the generic (most probable) configuration that none of them
lies in the tangent space of the torus. In the opposite case, the results do
not change qualitatively since GALI$_k$ still exhibit power law decays, but
with slightly different exponents \cite{SBA_07}. Defining by $\mbox{\boldmath
$\xi$}_i^{k}$ and $\mbox{\boldmath $\eta$}_i^{k}$ the $k\times 1$ column
matrices of initial conditions, the matrix $\textbf{D}$ assumes the form
%\begin{widetext}
\begin{equation}
\textbf{D}(t) \sim \frac{1}{t^{k}} \cdot \textbf{D}^{k}(t) = \frac{1}{t^{k}}
\left[
\begin{array}{ccccccccc}
\mbox{\boldmath $\xi$}_1^{k} & \ldots & \mbox{\boldmath $\xi$}_N^{k} &
\mbox{\boldmath $\eta$}_1^{k}+\sum_{i=1}^{s} \omega_{1i} \mbox{\boldmath
$\xi$}_i^{k}t & \ldots & \mbox{\boldmath $\eta$}_s^{k}+\sum_{i=1}^{s}
\omega_{si} \mbox{\boldmath $\xi$}_i^{k}t & \mbox{\boldmath $\eta$}_{s+1}^{k}
& \ldots & \mbox{\boldmath $\eta$}_{N}^{k}
\end{array} \right]  ,
\label{eq:matix _D2a}
\end{equation}
%\end{widetext}
where we have replaced the first factor on the right by its asymptotic
expression for long times.

In the case of an $s$--dimensional torus, the $k$ deviation vectors eventually
fall on its $s$--dimensional tangent space spanned by
$\hat{v}_{N+1},\ldots,\hat{v}_{N+s}$. If we start with $2\leq k \leq s$
deviation vectors, since there is no reason for them to become linearly
dependent, their wedge product yields GALI$_k$ indices that are different from
zero. However, if we start with $s< k \leq 2N$ deviation vectors, some of them
will necessarily become linearly dependent and thus their wedge product (as
well as the GALI$_k$) will tend to zero not exponentially but by {\it a power
law}, as we explain below.

In order to determine the leading order behavior of the GALI$_k$, we search
for the fastest increasing determinants of all $k \times k$ minors of the
matrix $\textbf{D}^{k}$, as $t$ grows. For $2\leq k \leq s$, these
determinants have $k$ columns chosen among the $s$ columns of matrix
$\textbf{D}^{k}$ with $\omega_{ij} \neq 0$ and grow as $t^{k}$, thus providing
constant terms to the GALI$_k$. All other determinants contain at least one
column from the $2N-s$ time independent columns of matrix $\textbf{D}^{k}$ and
introduce terms that grow {\it slower} than $t^{k}$, having ultimately no
bearing on the behavior of GALI$_k$(t). This yields the important result that
$\mbox{GALI}_k (t) \sim$ constant for $2 \leq k \leq s$.

Next, we turn to the case of $s < k \leq 2N-s$. The fastest growing
determinants are again those containing the $s$ columns of the matrix
$\textbf{D}^{k}$ with $\omega_{ij} \neq 0$.  The remaining $k-s$ columns are
chosen among the $2(N-s)$ columns of $\textbf{D}^{k}$ which are time
independent (excluding the $\mbox{\boldmath $\xi$}_i^{k}$ columns with $i \leq
s$). Among these determinants, the fastest increasing ones are those
containing as many columns proportional to $t$ as possible. Thus, $t$ appears
at most $s$ times and the time evolution of GALI$_k$ is mainly determined by
terms proportional to $t^{s}/t^k=1/t^{(k-s)}$ and hence $\mbox{GALI}_k (t)
\sim t^{(s-k)}$ for $s < k \leq 2N-s$.

Finally, let us consider the behavior of GALI$_k$ when $2N-s < k \leq
2N$. Again the fastest growing determinants contain the $s$ columns of
$\textbf{D}^{k}$ with $\omega_{ij} \neq 0$, while the rest $k-s$ columns are
chosen among the remaining $k-s$ time independent columns of $\textbf{D}^{k}$.
In order to have as many columns proportional to $t$ as possible these
determinants should contain $k-(2N-s)=k+s-2N$ columns among the
$\mbox{\boldmath $\xi$}_i^{k}$ columns with $i \leq s$, as well as the
corresponding $\mbox{\boldmath $\eta$}_i^{k}$ columns. Thus, $t$ appears at
most $s-(k+s-2N)=2N-k$ times and the time evolution of GALI$_k$ is determined
by terms proportional to $t^{2N-k}/t^k=1/t^{2(k-N)}$.  Summarizing, we have
shown that the GALI$_k$ for regular orbits lying on an $s$--dimensional torus
behave as \cite{Eleni_07}:
\begin{equation}
\mbox{GALI}_k (t) \sim \left\{ \begin{array}{ll} \mbox{constant} & \mbox{if
$2\leq k \leq s$} \\ \frac{1}{t^{k-s}} & \mbox{if $s< k \leq 2N-s$} \\
\frac{1}{t^{2(k-N)}} & \mbox{if $2N-s< k \leq 2N$} \\
\end{array}\right. .
\label{eq:GALI_order_all}
\end{equation}
Note that from (\ref{eq:GALI_order_all}) we deduce that for $s=N$, GALI$_k$
remains constant for $2\leq k \leq N$ and decreases to zero as $\sim
1/t^{2(k-N)}$ for $N< k \leq 2N$ in accordance with
(\ref{eq:GALI_order_all_N}).

%------------------------------------------------------
\section{Applications}
\label{Apl}

We now apply the GALI method to study chaotic, quasiperiodic and diffusive
motion in multi--dimensional Hamiltonian systems.  In particular, we consider
the FPU $\beta$--lattice of $N$ particles with Hamiltonian
\cite{Flach_05,Ford}
\begin{equation}
H = \sum_{i=1}^{N} \frac{p_i^2}{2} + \sum_{i=0}^{N}
\left[\frac{(q_{i+1}-q_i)^2}{2} + \frac{\beta (q_{i+1}-q_i)^4}{4} \right] \,
,\label{eq:FPUHam}
\end{equation}
with $q_1,\ldots, q_N$ being the displacements of the particles with respect
to their equilibrium positions, and $p_1,\ldots, p_N$ the corresponding
momenta. It is well known that if we define normal mode variables by
\begin{equation}
Q_k = \sqrt{ \frac{2}{N+1} }\sum_{i=1}^{N}
q_i\sin\left(\frac{ki\pi}{N+1}\right)\,\, ,\,\, P_k = \sqrt{ \frac{2}{N+1}
}\sum_{i=1}^{N} p_i\sin\left(\frac{ki\pi}{N+1}\right)\,\, ,\,\, k=1, \ldots,
N,
\label{eq:PQ}
\end{equation}
the unperturbed Hamiltonian (Eq.~(\ref{eq:FPUHam}) for $\beta=0$) is written
as the sum of the so--called {\it harmonic energies} $E_i$ having the form:
\begin{equation}
E_i=\frac{1}{2}\left( P_i^2 +\omega_i^2 Q_i^2\right) \,\, ,\,\, \omega_i=2
\sin \left( \frac{i\pi}{2(N+1)} \right) i=1, \ldots, N,
\label{eq:harm_e}
\end{equation}
with $\omega_i$ being the corresponding {\it harmonic frequencies}.  In our
study we impose fixed boundary conditions
$q_0(t)=q_{N+1}(t)=p_0(t)=p_{N+1}(t)=0$, $\forall t$ and fix the number of
particles to $N=8$ and the system's parameter to $\beta=1.5$.

\begin{figure}
\centerline{\includegraphics[scale=0.7]{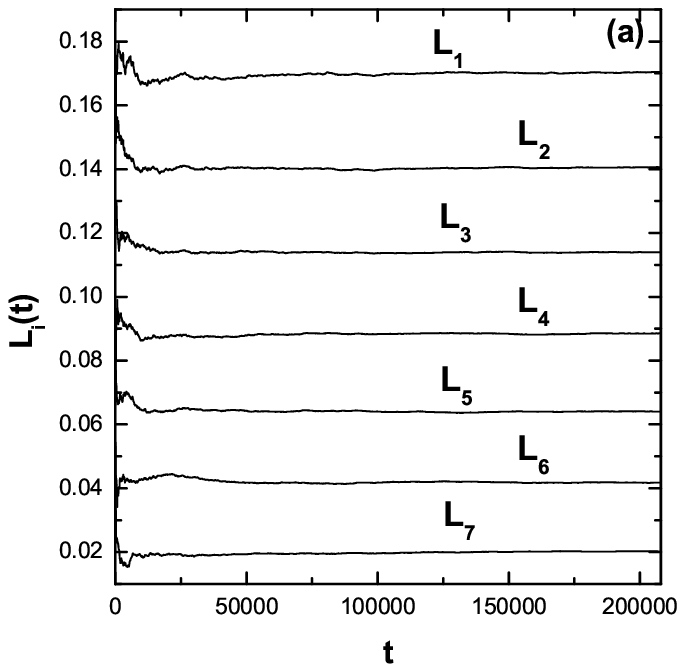} \hspace{-0.5cm}
\includegraphics[scale=0.7]{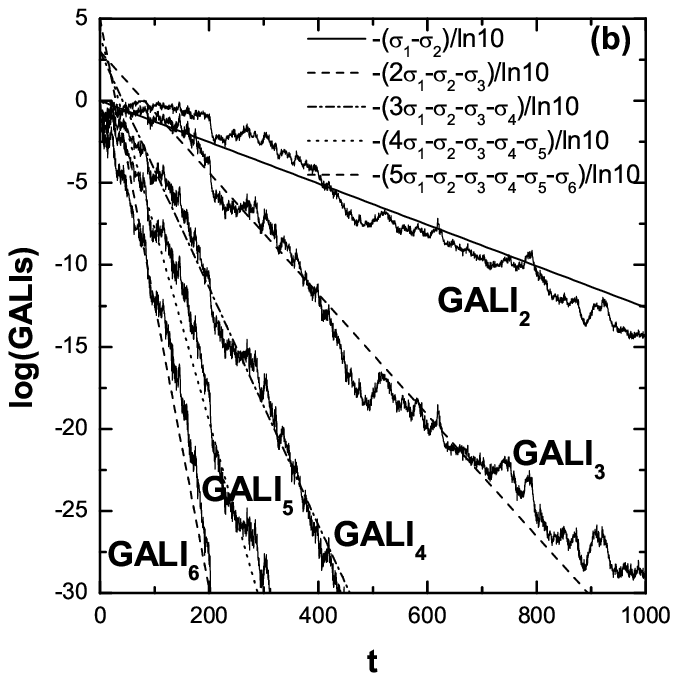} \hspace{-0.5cm}
\includegraphics[scale=0.7]{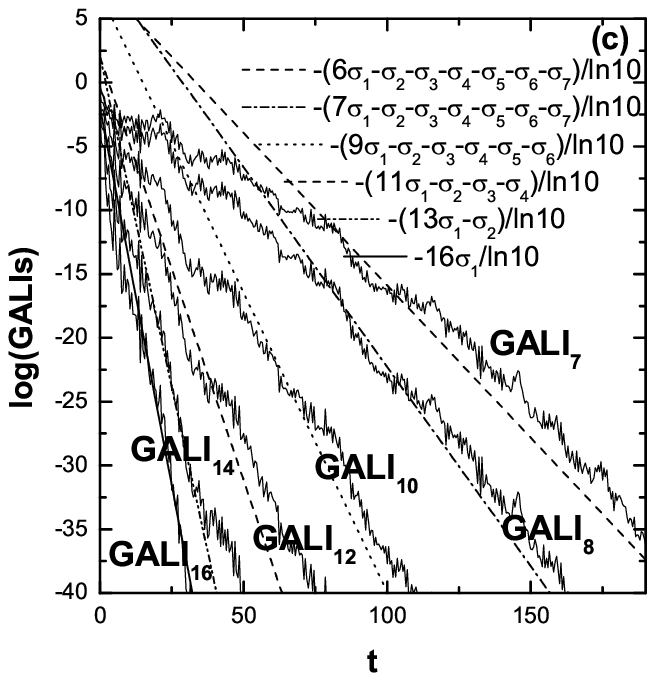}} \caption{(a) The time evolution of
quantities $L_i$, $i=1,\ldots,7$, having as limits for $t \rightarrow \infty$
the seven positive Lyapunov exponents $\sigma_i$, $i=1,\ldots,7$, for a
chaotic orbit with initial conditions $Q_1=Q_4=2$, $Q_2=Q_5=1$, $Q_3=Q_6=0.5$,
$Q_7=Q_8=0.1$, $P_i=0$, $i=1,\ldots,8$ of the $N=8$ particle FPU lattice
(\ref{eq:FPUHam}). The time evolution of the corresponding GALI$_k$ is plotted
in (b) for $k=2,\ldots,6$ and in (c) for $k=7,8,10,12,14,16$. The plotted
lines in (b) and (c) correspond to exponentials that follow the asymptotic
laws (\ref{eq:GALI_chaos}) for $\sigma_1=0.170$, $\sigma_2=0.141$,
$\sigma_3=0.114$, $\sigma_4=0.089$, $\sigma_5=0.064$, $\sigma_6=0.042$,
$\sigma_7=0.020$. Note that $t$--axis is linear and that the slope of each
line is written explicitly in (b) and (c).}
\label{fig:ch_GALIs}
\end{figure}

We consider first a chaotic orbit of (\ref{eq:FPUHam}), having seven positive
Lyapunov exponents, which we compute as the limits for $t \rightarrow \infty$
of some appropriate quantities $L_i$, $i=1,\ldots,7$ (see \cite{BGGS80b} for
more details), to be $\sigma_1\approx0.170$, $\sigma_2\approx0.141$,
$\sigma_3\approx0.114$, $\sigma_4\approx0.089$, $\sigma_5\approx0.064$,
$\sigma_6\approx0.042$, $\sigma_7\approx0.020$ (Fig.~\ref{fig:ch_GALIs}(a)).
We recall that chaotic orbits of Hamiltonian systems possess Lyapunov
exponents which are real and grouped in pairs of opposite sign, with two of
them being equal to zero. We, therefore, have in the case of the $N=8$
particle FPU lattice (\ref{eq:FPUHam}) $\sigma_i=-\sigma_{17-i}$ for
$i=1,\ldots,8$ with $\sigma_8=\sigma_9=0$. Using the above computed values as
good approximations of the real Lyapunov exponents, we see in
Figs.~\ref{fig:ch_GALIs}(b) and \ref{fig:ch_GALIs}(c) that the slopes of all
GALI$_k$ indices are well reproduced by (\ref{eq:GALI_chaos}).
%&&&&&&&&&&&&&&&&&&&&&&&&&&&&&&&&&&&&&&&&&
%&&&&&&&&&&&&&&&&&&&&&&&&&&&&&&&&&&&&&&&&&

\begin{figure}
\centerline{\includegraphics[scale=0.7]{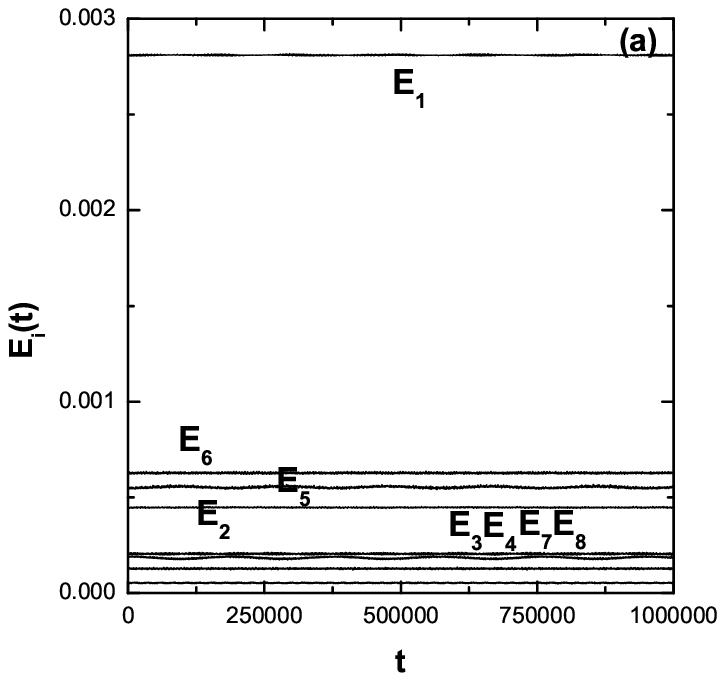} \hspace{-0.5cm}
\includegraphics[scale=0.7]{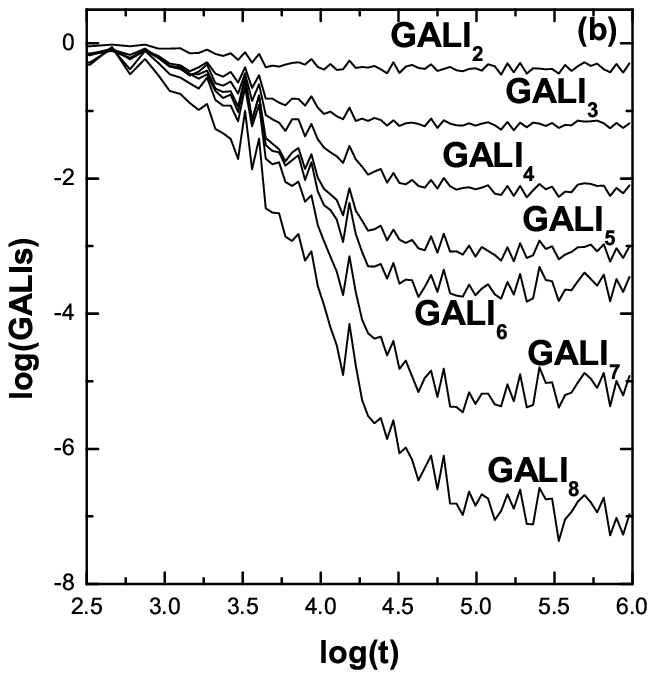} \hspace{-0.5cm}
\includegraphics[scale=0.7]{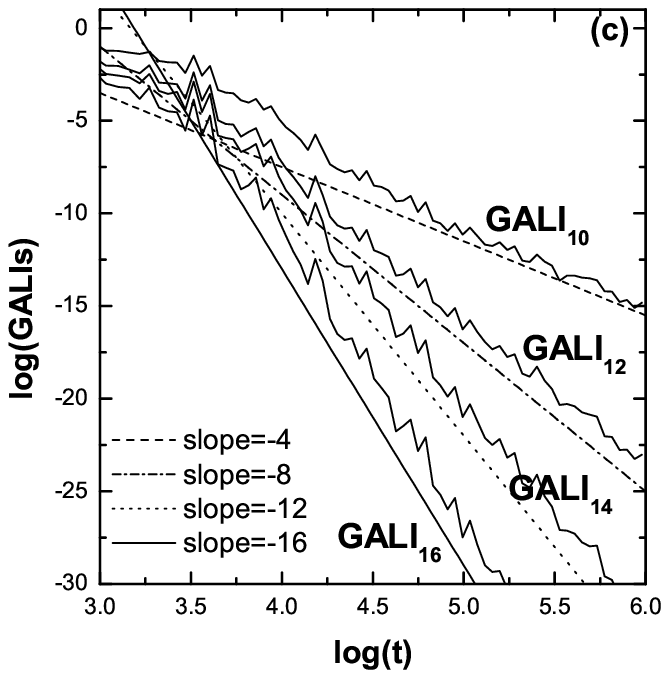}} \caption{(a) The time evolution of
harmonic energies $E_i$, $i=1,\ldots,8$, for a regular orbit with initial
conditions $q_1=q_2=q_3=q_8=0.05$, $q_4=q_5=q_6=q_7=0.1$, $p_i=0$,
$i=1,\ldots,8$ of the $N=8$ particle FPU lattice (\ref{eq:FPUHam}).  The time
evolution of the corresponding GALI$_k$ is plotted in (b) for $k=2,\ldots,8$
and in (c) for $k=10,12,14,16$. The plotted lines in (c) correspond to
functions proportional to $t^{-4}$, $t^{-8}$, $t^{-12}$ and $t^{-16}$, as
predicted in (\ref{eq:GALI_order_all_N}).}
\label{fig:reg_GALIs}
\end{figure}

Turning now to the case of a regular orbit of (\ref{eq:FPUHam}), we plot in
Fig.~\ref{fig:reg_GALIs}(a) the evolution of its harmonic energies $E_i$,
$i=1,\ldots,8$. The harmonic energies remain practically constant, exhibiting
some feeble oscillations, implying the regular nature of the orbit. In
Figs.~\ref{fig:reg_GALIs}(b) and \ref{fig:reg_GALIs}(c) we plot the GALIs of
this orbit and verify that their behavior is well approximated by the
asymptotic formula (\ref{eq:GALI_order_all_N}) for $N=8$. Note that the
GALI$_k$ for $k=2,\ldots,8$ (Fig.~\ref{fig:reg_GALIs}(b)) remain different
from zero implying that the orbit is indeed quasiperiodic and lies on a
8--dimensional torus. In particular, after some initial transient time, they
start oscillating around non--zero values whose magnitude decreases with
increasing $k$. On the other hand, the GALI$_k$ with $8 < k \leq 16$
(Fig.~\ref{fig:reg_GALIs}(c)) tend to zero following power law decays in
accordance to (\ref{eq:GALI_order_all_N}).
%&&&&&&&&&&&&&&&&&&&&&&&&&&&&&&&&&&&&&&&&&
%&&&&&&&&&&&&&&&&&&&&&&&&&&&&&&&&&&&&&&&&&

From the results of Figs.~\ref{fig:ch_GALIs} and \ref{fig:reg_GALIs}, we
conclude that the different behavior of GALI$_k$ for chaotic (exponential
decay) and regular orbits (non--zero values or power law decay) allows for a
fast and clear discrimination between the two cases. Let us consider for
example GALI$_8$ which tends exponentially to zero for chaotic orbits
(Fig.~\ref{fig:ch_GALIs}(c)), while it remains small but different from zero
in the case of regular orbits (Fig.~\ref{fig:reg_GALIs}(b)). At $ t \approx
150$ GALI$_8$ has values that differ almost 35 orders of magnitude being
GALI$_8 \approx 10^{-36}$ for the chaotic orbit, while GALI$_8 \approx
10^{-1}$ for the regular one. This huge difference in the values of GALI$_8$
clearly identifies the chaotic nature of the orbit.

If we select initial conditions such that only a small number of the harmonic
energies $E_i$ are initially excited, we observe, at small enough energies
that (\ref{eq:FPUHam}) exhibits the famous FPU recurrences, whereby energy is
exchanged quasiperiodically only between the excited $E_i$s.

\begin{figure}
\centerline{\includegraphics[scale=0.7]{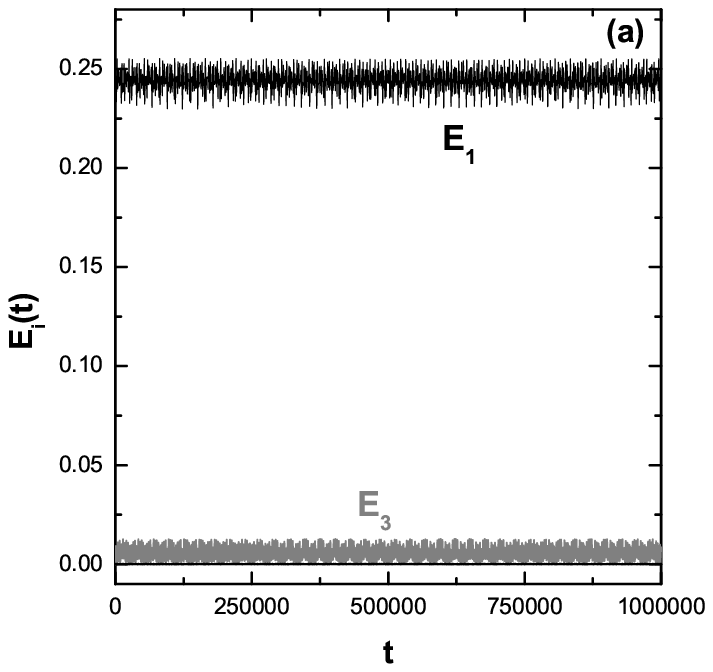} \hspace{-0.5cm}
\includegraphics[scale=0.7]{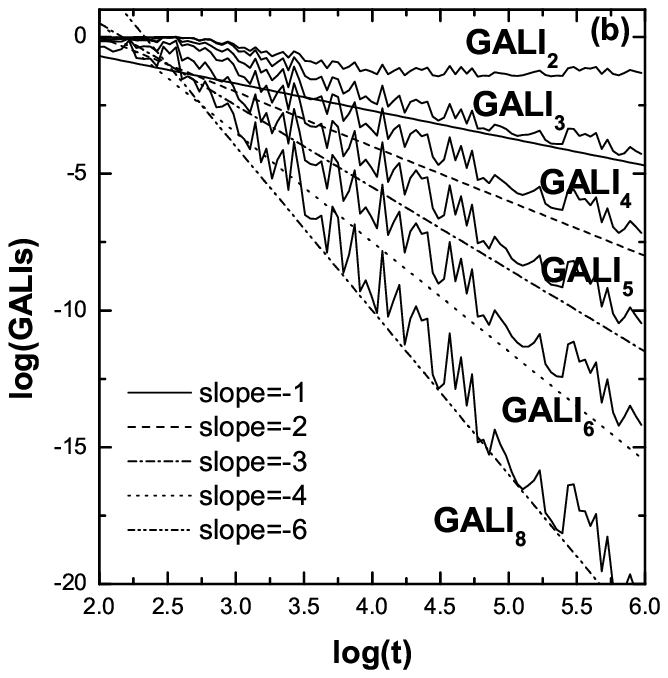} \hspace{-0.5cm}
\includegraphics[scale=0.7]{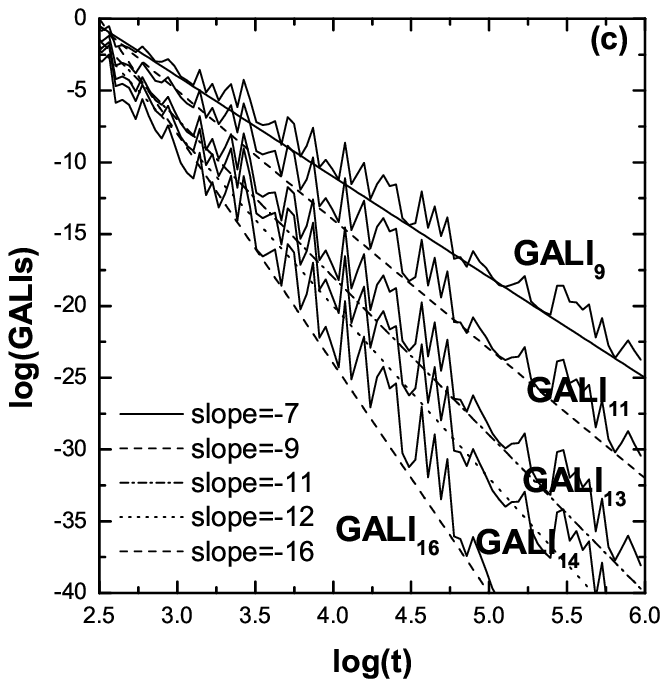}} \caption{(a) The time evolution of
harmonic energies for a regular orbit lying on a 2--dimensional torus of the
$N=8$ particle FPU lattice (\ref{eq:FPUHam}). Recurrences occur between $E_1$
and $E_3$, while all other harmonic energies remain practically zero.  The
time evolution of the corresponding GALI$_k$ is plotted in (b) for
$k=2,\ldots,6,8$ and in (c) for $k=9,11,13,14,16$. The plotted lines in (b)
and (c) correspond to precisely the power laws predicted in
(\ref{eq:GALI_order_all}).} \label{fig:FPU_2d}
\end{figure}

In Fig.~\ref{fig:FPU_2d}, we consider the case of a regular orbit with initial
conditions $Q_1=2$, $P_1=0$, $Q_i=P_i=0$, $i=2,\ldots,8$ having total energy
$H=0.24$.  In Fig.~\ref{fig:FPU_2d}(a) we see that only two normal modes are
exited, namely $E_1$ and $E_3$, while all other harmonic energies remain
practically zero.  Observe that among the GALI$_k$ plotted in
Figs.~\ref{fig:FPU_2d}(b) and (c), only GALI$_2$ $\approx$ const., indicating
that the torus is only 2--dimensional, while all others decay by power laws
whose exponents are the ones given in (\ref{eq:GALI_order_all}) for $N=8$ and
$s=2$ (see also \cite{Eleni_07} for more details).
%&&&&&&&&&&&&&&&&&&&&&&&&&&&&&&&&&&&&&&&&&
%&&&&&&&&&&&&&&&&&&&&&&&&&&&&&&&&&&&&&&&&&

\begin{figure}
\centerline{\includegraphics[scale=0.7]{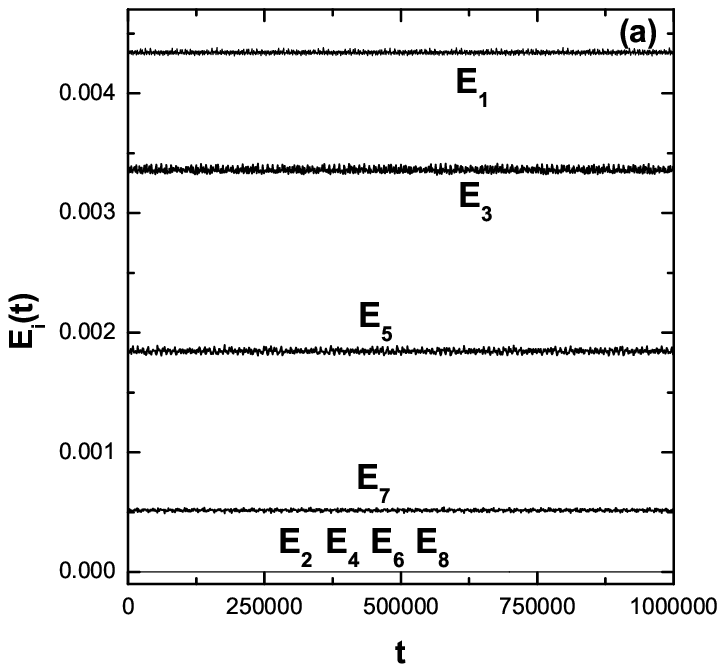} \hspace{-0.5cm}
\includegraphics[scale=0.7]{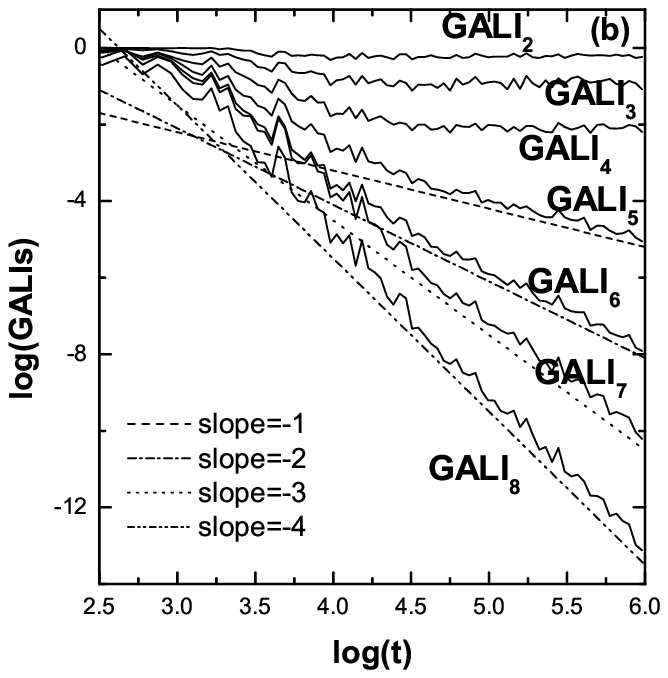} \hspace{-0.5cm}
\includegraphics[scale=0.7]{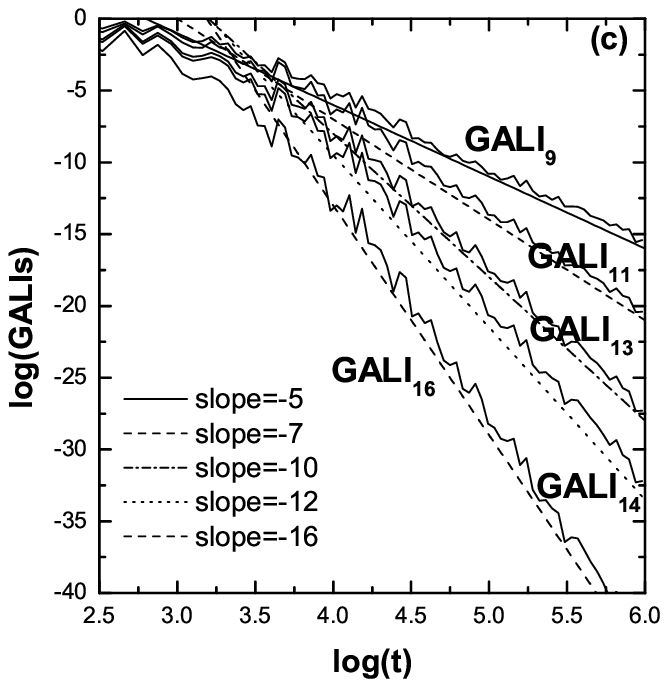}} \caption{(a) The time evolution of
harmonic energies for a regular orbit lying on a 4--dimensional torus of the
$N=8$ particle FPU lattice (\ref{eq:FPUHam}). Recurrences occur between $E_1$,
$E_3$, $E_5$ and $E_7$, while all other harmonic energies remain practically
zero.  The time evolution of the corresponding GALI$_k$ is plotted in (b) for
$k=2,\ldots,8$ and in (c) for $k=9,11,13,14,16$. The plotted lines in (b) and
(c) correspond to the precise power laws predicted in
(\ref{eq:GALI_order_all}).} \label{fig:FPU_4d}
\end{figure}

Let us now choose initial conditions so as to distribute the energy among 4
modes, $E_i$, $i=1,3,5,7$, in our 8--particle FPU lattice.  In particular, we
consider an orbit with initial conditions $q_i=0.1$, $p_i=0$, $i=1,\ldots,8$,
having total energy $H=0.01$.  What we observe again is that since only these
4 modes are excited (Fig.~\ref{fig:FPU_4d}(a)), only the GALI$_k$ for
$k=2,3,4$ remain constant (Fig.  \ref{fig:FPU_4d}(b)), implying that the
motion lies on a 4--dimensional torus, while all the higher order GALIs decay
by power laws, whose exponents are derived in (\ref{eq:GALI_order_all}) for
$N=8$ and $s=4$.
%&&&&&&&&&&&&&&&&&&&&&&&&&&&&&&&&&&&&&&&&&
%&&&&&&&&&&&&&&&&&&&&&&&&&&&&&&&&&&&&&&&&&

\begin{figure}
\centerline{\includegraphics[scale=0.7]{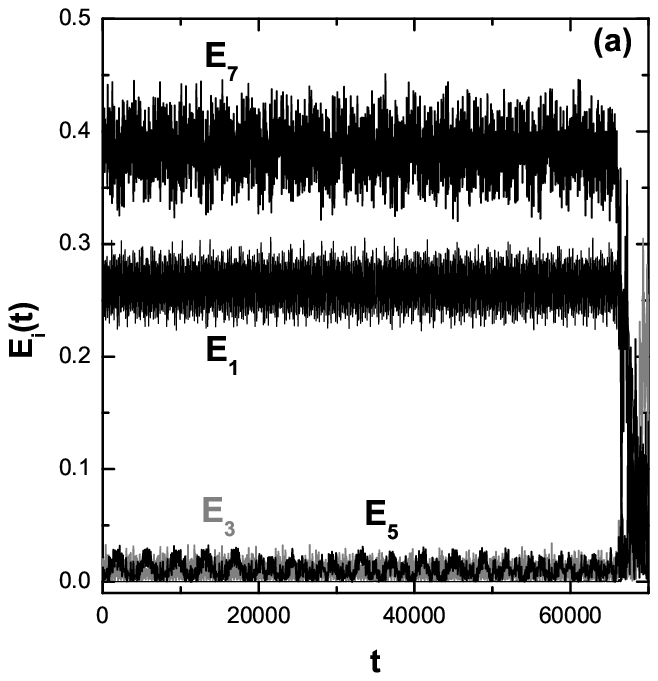} \hspace{-0.5cm}
\includegraphics[scale=0.7]{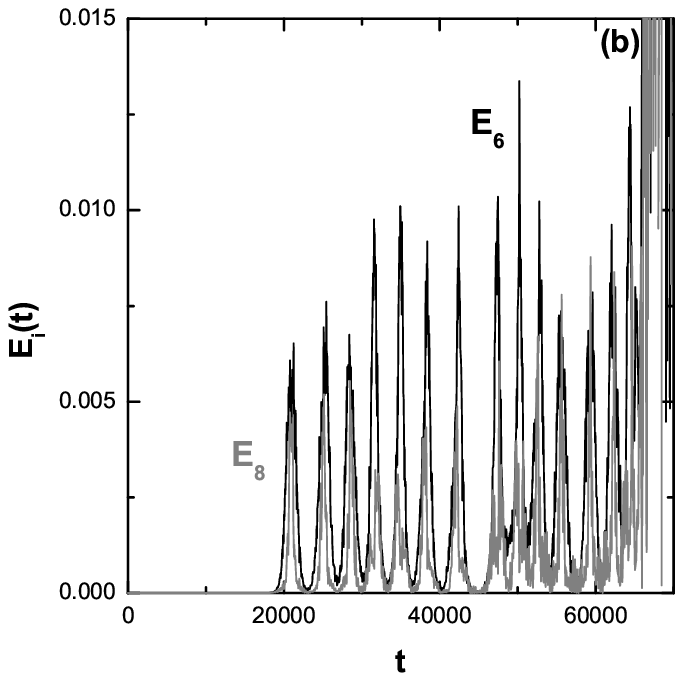} \hspace{-0.5cm}
\includegraphics[scale=0.7]{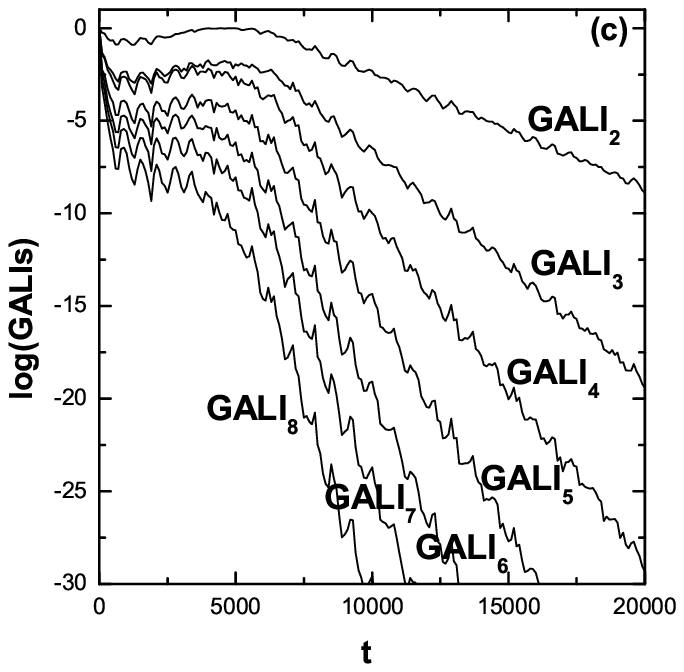}} \caption{The time evolution of
harmonic energies (a) $E_1$, $E_3$, $E_5$, $E_7$ and (b) $E_6$, $E_8$, for a
slowly diffusing orbit of the $N=8$ particle FPU lattice
(\ref{eq:FPUHam}). (c) The time evolution of the corresponding GALI$_k$ for
$k=2,\ldots,8$, clearly exhibits exponential decay already at $t\approx
10000$.} \label{fig:dif}
\end{figure}

What happens, however, if we choose an orbit that starts near a torus but
slowly drifts away from it, presumably through a thin chaotic layer of higher
order resonances?  This phenomenon is recognized by the GALIs, which provide
early predictions that may be quite relevant for applications. To see this let
us choose again initial conditions for our 8--particle FPU lattice, such that
the motion appears quasiperiodic, with its energy recurring between the modes
$E_1$, $E_3$, $E_5$ and $E_7$ (Fig.~\ref{fig:dif}(a)). In particular, we
consider an orbit with initial conditions $Q_1=2$, $Q_7=0.44$,
$Q_3=Q_4=Q_5=Q_6=Q_8=0$, $P_i=0$, $i=1,\ldots,8$, having total energy
$H=0.71$.  This orbit, however, is {\it not} quasiperiodic, as it drifts away
from the initial 4--dimensional torus, exciting new frequencies and sharing
its energy with more modes, after about $t=20000$ time units. This becomes
evident in Fig.~\ref{fig:dif}(b) where we plot the evolution of $E_6$ and
$E_8$. We see that these harmonic energies, which were initially zero, start
having non--zero values at $t \approx 20000$ and exhibit from then on small
oscillations (note the different scales of ordinate axis of
Figs.~\ref{fig:dif}(a) and (b)), which look very regular until $t \approx
66000$. At that time the values of all harmonic energies change dramatically,
clearly indicating the chaotic nature of the orbit. As we see in
Fig.~\ref{fig:dif}(c), this type of diffusion is predicted by the {\it
exponential} decay of {\it all} GALI$_k$, shown already at about $t=10000$.
%&&&&&&&&&&&&&&&&&&&&&&&&&&&&&&&&&&&&&&&&&
%&&&&&&&&&&&&&&&&&&&&&&&&&&&&&&&&&&&&&&&&&

%------------------------------------------------------
\section{Summary}
\label{sum}

We have applied the Generalized Alignment Index of order $k$ (GALI$_k$) as a
tool for studying the dynamics in conservative dynamical systems, and in
particular in Hamiltonian systems of $N$ degrees of freedom. We have shown
that these indices not only distinguish rapidly between chaotic and regular
orbits, but also determine the dimensionality of quasiperiodic tori and detect
slow diffusion away from quasiperiodicity, long before this becomes evident in
the dynamics.

The GALI$_k$ represents the `volume' of a generalized parallelogram having as
edges $k>2$ initially linearly independent unit deviation vectors and are
computed as the norm of the wedge product of these vectors. We verified
numerically that for chaotic orbits GALI$_k$ tends exponentially to zero
following a rate which depends on the values of several Lyapunov exponents
(see Eq.~(\ref{eq:GALI_chaos})). In the case of regular orbits lying on
$N$--dimensional tori, GALI$_k$ with $2 \leq k \leq N$ eventually fluctuates
around non--zero values, while for $N< k\leq 2N$, it tends to zero following a
particular power law (see Eq.~(\ref{eq:GALI_order_all_N})). If, on the other
hand, the orbit lies on an $s$--dimensional torus, with $1\leq s\leq N$, we
showed analytically and verified numerically that GALI$_k$ is nearly constant
for $2 \leq k \leq s$ and, if $s<k\leq 2N$, decays as a power law, the precise
form of which depends on the dimensionality of the torus (see
Eq.~(\ref{eq:GALI_order_all})).

Computationally, of course, it is somewhat costly to evaluate the large number
of determinants needed to obtain the GALI$_k$s by Eq.~(\ref{eq:norm}),
especially in the case of large $N$. As a solution to this problem, we
introduced and theoretically explained in Sect.~\ref{GALI-num} an efficient
and accurate method of computing $\log (\mbox{GALI}_k)$ by applying the
technique of Singular Value Decomposition (SVD) to the matrix of deviation
vectors and evaluating the product of its singular values (see
Eq.~(\ref{eq:gali_svd})).

Applying next the GALI method to a 1--dimensional Fermi--Pasta--Ulam (FPU)
lattice of $N=8$ particles, we have demonstrated that GALIs do answer
efficiently and reliably some fundamental questions of practical concern: When
is the motion quasiperiodic and what is the dimensionality of the torus on
which it lies? This is quite important when one wishes to determine the number
of frequencies involved and leads to the interesting result that the dimension
can be much less than the generally expected number of degrees of freedom
$N$. Also, when is the motion {\it not quasiperiodic} but diffuses slowly away
from a torus through a chaotic network of higher order resonances? This is
notoriously difficult to ascertain (especially for large $N$), as the dynamics
is `sticky', Lyapunov exponents may be very small and thus long integrations
are needed before the chaotic nature of the motion becomes evident.

We believe that the results we have presented on the FPU lattice demonstrate
that the GALI$_k$ method can prove very useful to many practical applications:
In celestial mechanics, there are many higher dimensional $N$--body problems
\cite{Contopoulos,MeBe_07}, where one would like to locate invariant tori and
determine the extent of phase space occupied by regular and chaotic orbits. In
chemical dynamics, small molecules have been recently studied whose properties
depend on the presence of stable oscillatory modes carrying large regions of
quasiperiodic motion about them \cite{Jung_06}. Many interesting problems are
also described by $N$-dimensional symplectic maps, as e.g. in accelerator
dynamics, where the presence of large regions of tori is essential for
maximizing the stability of betatron oscillations \cite{Scandale}.

%-------------------------------------------------
\begin{acknowledgement}
Ch.~S. was supported by the Marie Curie Intra--European Fellowship No
MEIF--CT--2006--025678. This work was partially supported by the European
Social Fund (ESF), Operational Program for Educational and Vocational Training
II (EPEAEK II) and particularly the Program PYTHAGORAS II. We would also like
to thank both referees for their remarks.
\end{acknowledgement}

%-------------------------------------------------


\begin{thebibliography}{}

\bibitem{Chirikov} B.~V.~Chirikov, Phys.~Rep. {\bf
180}, 179 (1979)

\bibitem{MacKay_1987} R.~S.~MacKay and J.~D.~Meiss, {\it Hamiltonian Dynamical
Systems}, (Adam Hilger, Bristol, 1987)

\bibitem{LL92} M.~A.~Lieberman and A.~J.~Lichtenberg, {\it Regular and Chaotic
Dynamics} (Springer Verlag, 1992)

\bibitem{Contopoulos} G.~Contopoulos, {\it Order and Chaos in Dynamical
Astronomy}, (Springer, Berlin, 2002)

\bibitem{Scandale} W.~Scandale and G.~Turchetti, eds., {\it Nonlinear Problems
in Future Particle Accelerators}, (World Scientific, Singapore, 1991);
T.~Bountis and Ch.~Skokos, Nucl. Instr. Meth. Res. A {\bf 561}, 173 (2006) and
other articles in that volume

\bibitem{Farantos} S.~C.~Farantos, Z--W. Qu, H.~Zhu and R.~Scinke,
Int. J. Bif. Chaos., {\bf 16} (7), 1913 (2006)

\bibitem{Jung_06} C.~Jung, H.~S.~Taylor and E.~L.~Sibert, J.~Phys.~Chem.~A
{\bf 110}, 5317 (2006)

\bibitem{Flach_05} S.~Flach, M.~V.~Ivanchenko and O.~I.~Kanakov, PRL {\bf 95},
  064102 (2005)

\bibitem{Ford} J.~Ford, Phys.~Rep., {\bf 213}, 271 (1992); G.~P.~Berman and
F.~M.~Izrailev, Chaos {\bf 15}, 015104 (2005)

\bibitem{BGGS80a} G.~Benettin, L.~Galgani, A.~Giorgilli and J.--M.~Strelcyn,
Meccanica, {\bf March}, 9 (1980)

\bibitem{BGGS80b} G.~Benettin, L.~Galgani, A.~Giorgilli and J.--M.~Strelcyn,
Meccanica, {\bf March}, 21 (1980)

\bibitem{SBA_07} Ch.~Skokos, T.~C.~Bountis and Ch.~Antonopoulos, Physica D
{\bf 231}, 30 (2007)

\bibitem{S01} Ch.~Skokos, J. Phys. A {\bf 34}, 10029 (2001); Ch.~Skokos,
Ch.~Antonopoulos, T.~C.~Bountis and M.~N.~Vrahatis, Prog. Theor. Phys. Supp.,
{\bf 150} 439 (2003); Ch.~Skokos, Ch.~Antonopoulos, T.~C.~Bountis and
M.~N.~Vrahatis, J. Phys. A {\bf 37}, 6269 (2004)

\bibitem{Sticky} F.~F.~Karney, Physica D {\bf 8}, 360 (1983); J.~D.Meiss and
E.~Ott, Phys.~Rev.~Let. {\bf 55}, 2741 (1985); V.~Afraimovich and
G.~M.~Zaslavsky, Lect.~Notes Phys., {\bf 511}, 59 (1998); Ch.~Efthymiopoulos,
G.~Contopoulos, N.~Voglis and R.~Dvorak, J. Phys. A {\bf 30}, 8167 (1997);
R.~Dvorak, G.~Contopoulos, Ch.~Efthymiopoulos and N.~Voglis, Planet. Space
Sci., {\bf 46}, 1567 (1998)

\bibitem{Spivak_1999} M.~Spivak, {\it Comprehensive Introduction to
Differential Geometry}, vol.~1, (Publ. or Per. Inc., 1999)

\bibitem{AThesis} Ch.~Antonopoulos and T.~Bountis, ROMAI Journal {\bf 2} (2),
1 (2006); see also Ch.~Antonopoulos Ph.~D.~Thesis, Department of Mathematics,
University of Patras (2007)

\bibitem{HH} J.~H.~Hubbard and B.~B.~Hubbard, {\it Vector Calculus, Linear
Algebra and Differential Forms: A Unified Approach}, Chapter 5, (Prentice
Hall, 1999)

\bibitem{B58} N.~Bourbaki, {\it \'{E}l\'{e}ments de math\'{e}matique, Livre
II: Alg\`{e}bre}, Chapitre 3, (Hermann, 1958)

\bibitem{NumRec} W.~H.~Press, S.~A.~Teukolsky, W.~T.~Vetterling and
B.~P.~Flannery, {\it Numerical Recipes in Fortran 77: The Art of Scientific
Computing}, Chapter 2, (Cambridge University Press, 2003).

\bibitem{MQ06} R.~I.~McLachlan and G.~R.~W.~Quispel, J. Phys. A {\bf 39}, 5251
(2006)

\bibitem{PD81} P.~J.~Prince and J.~R.~Dormand, J.~Comp.~Appl.~Math., {\bf 7},
67 (1981)


\bibitem{Eleni_07} E.~Christodoulidi and T.~Bountis, ROMAI Journal {\bf 2}
(2), 37 (2006)


\bibitem{MeBe_07} O.~Merlo and L.~Benet, Cel. Mech. Dyn.  Astr.  {\bf 97}, 49
(2007)

\end{thebibliography}
\end{document}